\documentclass[12pt]{iopart}

\usepackage{graphicx}
\usepackage{dcolumn}
\usepackage{bm}

\include{epsf} 

\begin{document}
\title{Indo-European languages tree by Levenshtein distance}

\author{Maurizio Serva}
\address{Dipartimento di Matematica,
Universit\`a dell'Aquila, I-67010 L'Aquila, Italy}
\ead{serva@univaq.it}

\author{Filippo Petroni}
\address{Dipartimento di Matematica,
Universit\`a dell'Aquila, I-67010 L'Aquila, Italy}
\address{GRAPES, B5,
Sart Tilman, B-4000 Li\`ege, Belgium}
\ead{filippo.petroni@ulg.ac.be}

\begin{abstract}
The evolution of languages closely resembles the evolution of
haploid organisms. 
This similarity has been recently exploited \cite{GA,GJ}
to construct language trees.
The key point is the definition of a distance among
all pairs of languages which is the analogous of a genetic distance.  
Many methods have been proposed to define these distances, one of this,
used by glottochronology, compute distance from 
the percentage of shared ``cognates''. 
Cognates are words inferred to have a common historical origin, and 
subjective judgment plays a relevant role in the identification process.
Here we push closer the analogy with evolutionary biology and we introduce
a genetic distance among language pairs by considering
a renormalized Levenshtein distance among words with same meaning
and averaging on all the words contained in a Swadesh list \cite{Sw}.
The subjectivity of process is consistently reduced and the
reproducibility is highly facilitated. \\
We test our method against the Indo-European group  considering fifty 
different languages and the two hundred words of the Swadesh 
list for any of them. We find out a tree which closely resembles the one published 
in \cite{GA} with some significant differences.
\end{abstract}

\section{Introduction}
Glottochronology uses the percentage of shared ``cognates'' between
languages to calculate their distances. These  ``genetic''  distances
are logarithmically proportional to divergence times 
if a constant rate of lexical replacement is assumed. 
Cognates are words inferred to have a common historical origin, 
their identification is often a matter of sensibility and personal
knowledge. Therefore, subjectivity plays a relevant role.
Furthermore,  results are often biased 
since it is easier for European or American scholars to find
out those cognates belonging to western languages.
For instance, the Spanish word {\it leche} and the Greek 
word {\it gala} are cognates.  In fact, {\it leche} comes
from the  Latin {\it lac} with genitive form {\it lactis},
while the genitive form of {\it gala} is {\it galactos}. 
This identification is possible
because of our historical records, hardly it would 
have been possible for languages, let's say, of Central Africa. 

Our aim is to avoid this subjectivity and construct a languages tree
which can be easily replicated by other scholars.
To reach this goal, we compare words with same meaning belonging to
different languages only considering orthographical differences.
More precisely, we use a modification of the Levenshtein distance (or edit distance)
to measure distance between pairs of words in different languages. 
The edit distance is defined as the minimum number of operations
needed to transform one word into the other, where an operation is an
insertion, deletion, or substitution of a single character.
Our definition of genetic distance between two words is taken as the edit distance
divided by the number of characters of the longer of the two. 
With this definition, the distance can take any value between 0 and 1.
To understand why we renormalize, let us consider the following case of
one substitution between two words: if the compared words are long even 
if the difference between them is given by one substitution they
remain very similar; while, if these words are short, let's say 
two characters, one substitution is enough to make them completely
different. Without renormalization, the distance between the 
words compared in the two examples would be the same, no matter their length. 
Instead, introducing the normalization factor, in the first case the
genetic distance would be much smaller than in the second one. 

We use the distance between words pairs, as defined above, to
construct a distance between pairs of languages. 
The first step is to find lists of words with the same meaning for all
the languages  for which we intend to construct the distance. 
Then, we compute the genetic distance for each pair of words 
with same meaning in one language pair. Finally, the distance between each
language pair is  defined as the average of the distance between words
pair.  As a result we have a number between 0 and 1 which we claim to
be the genetic distance between the two languages.

\section{Languages database}
The database we use for the present analysis \cite{footnote1} is composed by 50
languages with 200 words for each of them.
The words are chosen according to the Swadesh list. All the languages 
considered belong to the Indo-European group.
The database is a selection/modification of the one used in \cite{D},
where some errors have been corrected, and many missing words have been added.
In the database only the English alphabet is used (26 character plus
space); those languages written in a different alphabet 
(i.e. Greek etc.) were already transliterated into the English
one in \cite{D}.
For some of the languages in our lists \cite{footnote1}
there are still few missing words 
for a total number of 43 in a database of 9957. 
When a language has one or more missing words, 
these are simply not considered in the average that brings to the definition 
of distance. This implies that for some pairs of languages,  
the number of compared words is not 200 but a number
always greater than or equal to 187.
There is no bias in this procedure, the only effect
is that the statistic is slightly reduced.   

The result of the analysis described above is a $50  \times  50$
upper triangular matrix which expresses the 1225 distances among all languages pairs.

Indeed, our method for computing distances is a very simple operation, 
that does not need any specific linguistic knowledge and requires
a minimum computing time.

\section{Time distance between languages}
A phylogenetic tree can be build  already from this matrix,
but this would only give the topology of the tree
whereas the absolute time scale would be missing.
In order to  have this quantitative information, some hypotheses  
on the time evolution of genetic distances are necessary.
We assume that the genetic distance among words, on one side tends
to grow due to random mutations and on the other side may reduce
since different words may become more similar by accident or, 
more likely, by language borrowings.

Therefore, the  distance $D$ between two given languages can be thought to  
evolve according to the simple differential equation
\begin{equation}
\label{diffeq}
\dot{D}=-\alpha \,(1-D) -\beta D
\end{equation}
where $\dot{D}$ is the time derivative of $D$.
The parameter $\alpha$ is related to the increasing of $D$ 
due to random permutations, deletions or substitutions of characters 
(random mutations) while the parameter $\beta$ considers
the possibility that two words become more similar 
by a ``lucky'' random mutation or by words borrowing
from one language to the other or both from a third one. 
Since $\alpha$ and $\beta$ are constant, it is implicitly
assumed that mutations and borrowings occur at a constant rate.

At time $T=0$ the two languages begin
to separate and the genetic distance $D$ is zero.
With this initial condition the above equation can be solved
and the solution can be inverted. The result
is a relation which gives the separation time $T$ (time distance)
between two languages in terms of their genetic distance $D$
\begin{equation}
T= -\epsilon \,\ln(1 - \gamma D)
\label{time}
\end{equation}
The values for the parameters $\epsilon = 1/(\alpha + \beta)$
and  $\gamma = (\alpha + \beta ) / \alpha$ can be fixed 
experimentally by considering  two pairs of 
languages whose separation time (time distance) is known. 
We have chosen a distance of 1600 years between Italian and French
and a distance of 1100 years between Icelandic and Norwegian. 
The resulting values of the parameter are $\epsilon =1750$ and
$\gamma=1.09$ which corresponds to the following values $\alpha \cong 5*10^{-4}$ 
and $\beta\cong 6*10^{-5}$.
This means that similar words may become more different at a rate
that is about ten times the rate at which different words 
may become more similar. It should be noticed that (\ref{time})
closely resembles the fundamental formula of glottochronology.

The time distance $T$ is then computed for all pairs of languages 
in the database, obtaining a $50 \times  50 $ upper triangular with 
$1225$ non trivial entries. 
This matrix preserves the topology of the genetic distance matrix
but it contains all the information concerning absolute time scales.

\begin{figure}
\epsfysize=20.0truecm \epsfxsize=18.0truecm
\centerline{\epsffile{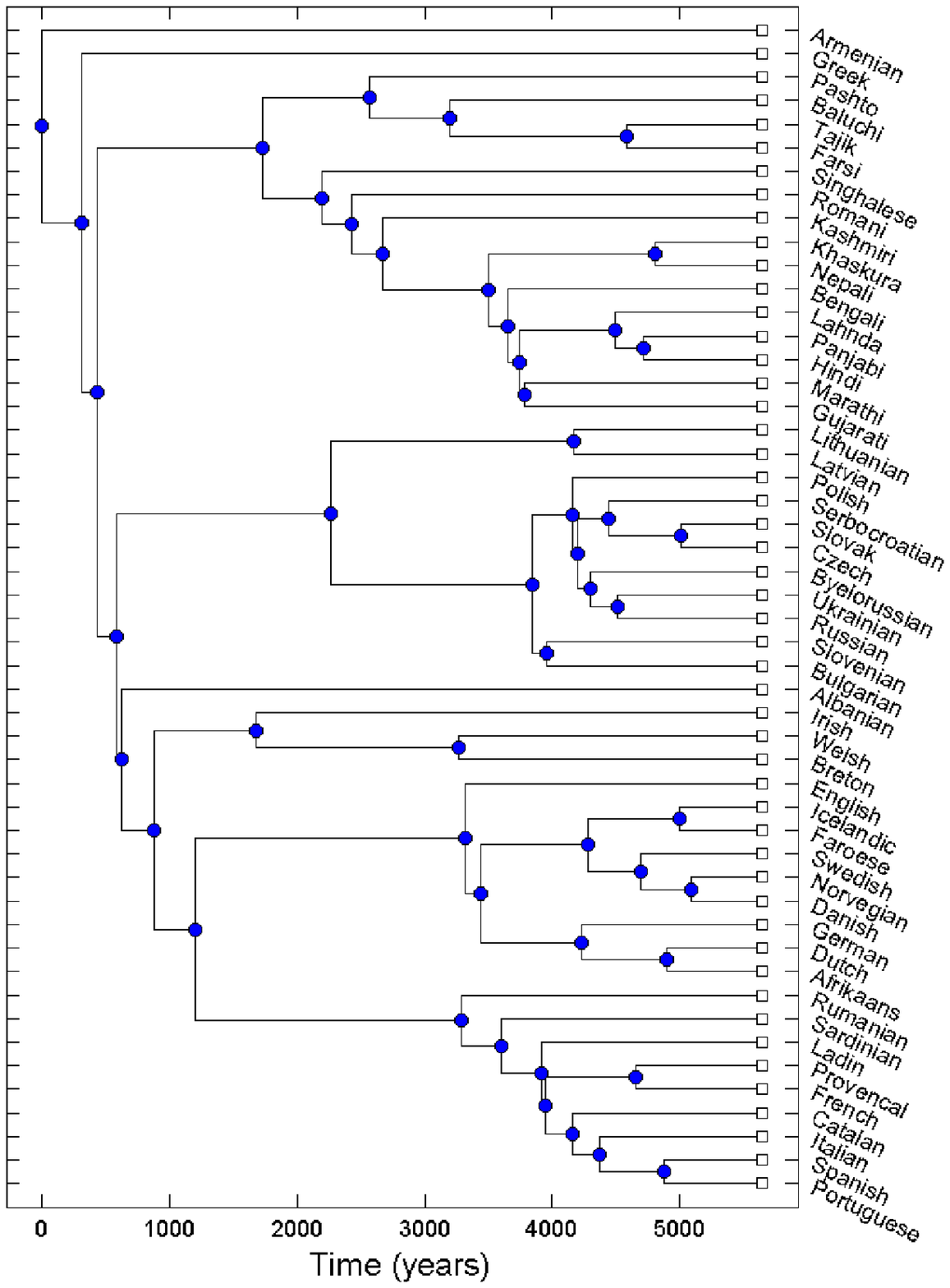}} \caption{}\label{fig1}
\end{figure}

\begin{figure}
\epsfysize=12.0truecm \epsfxsize=14.0truecm
\centerline{\epsffile{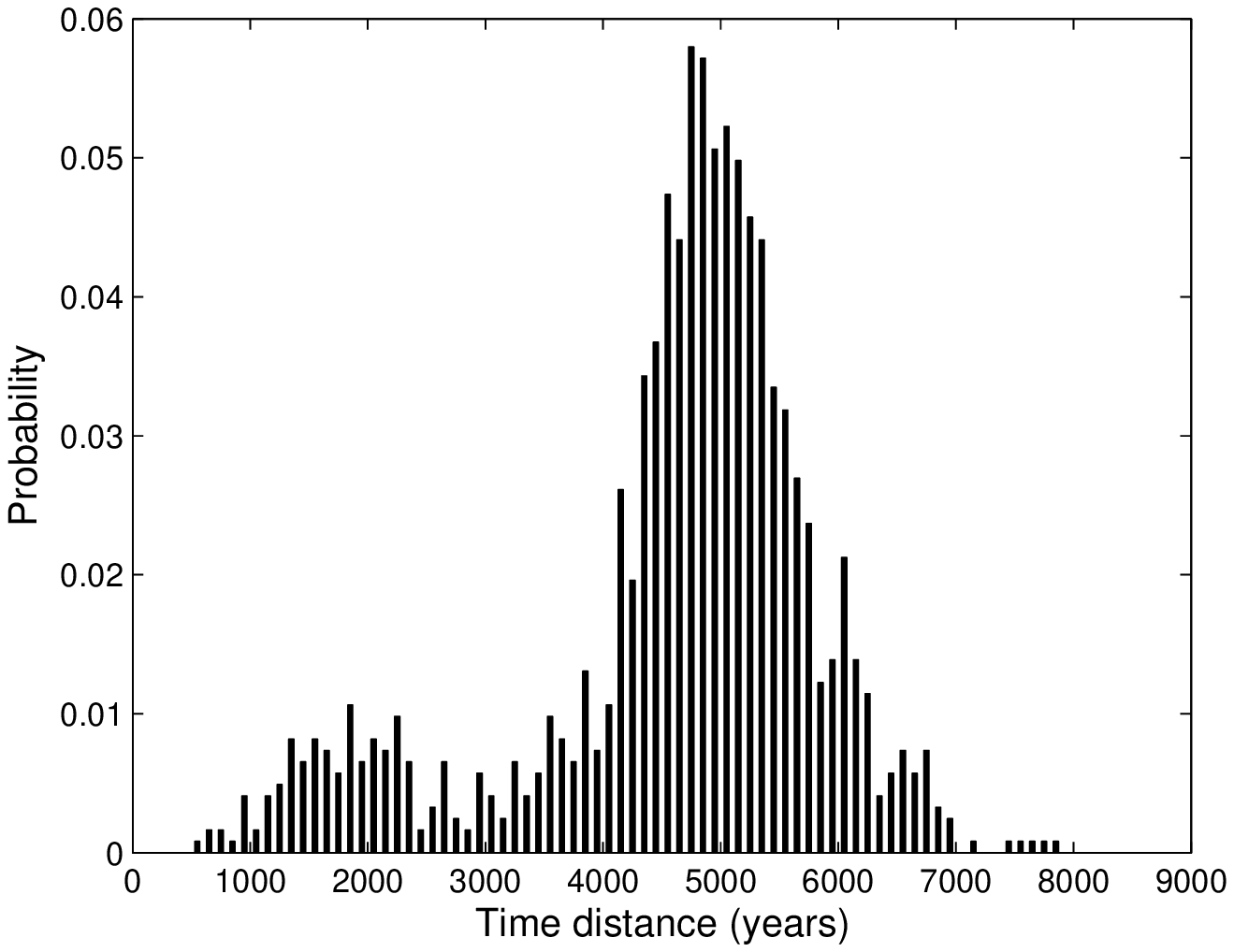}} \caption{}\label{fig2}
\end{figure}

The phylogenetic tree in Fig. \ref{fig1}  
is constructed from the matrix using the Unweighted Pair Group 
Method Average (UPGMA).
We use UPGMA for its coherence with the 
trees associated with the coalescence process of Kingamnn type \cite{K}.
In fact, the process of languages separation and extinction closely 
resembles the population dynamics associated with haploid reproduction 
which holds for simple organism or for the mitochondrial DNA 
of complex ones as humans.
This dynamics, introduced by Kingman, has been extensively
studied and described, see for example \cite{S,SD}.
In particular, in these two papers the distribution
of distances is found and plotted and can be usefully 
compared with the one herein obtained and plotted in Fig. \ref{fig2}.
It should be considered that in the model of Kingman,
time distances have the objective meaning of measuring the time
from separation while in our realistic case
time distances are reconstructed from genetic distances.
In this reconstruction we assume  that lexical mutations and
borrowings happen at a constant rate.
This is true only on average, since there is an 
inherent randomness in this process which is not taken into account by
the deterministic differential equation (\ref{diffeq}).
Furthermore, the parameters $\alpha$ and $\beta$  may vary from a pair of languages
to another and also they may vary in time 
according to historical conditions.
Therefore, the distribution in Fig. \ref{fig2}
is not exactly the distribution in \cite{S,SD} 
but it could be obtained from them after a random shift of all distances.

In this paper we do not consider dead languages,
we suspect, in fact,  that results would be biased due to the 
different times in which languages existed.
For example,  comparison of Latin with its offspring could be a 
meaningless operation in the context of this research.
We think that, eventually, Latin should be compared with its contemporary languages
and their genealogical tree constructed.
\section{Methods}
\subsection{Database}
The database used here to construct the phylogenetic tree is composed
by 50 languages of the Indo-European group. 
The main source for the database is the file prepared by Dyen et
al. in \cite{D} which contains the Swadesh list of 200 words for 96 languages.
This list consists of items of basic vocabulary, like body parts,
pronouns, numbers, which are known to be resistant to borrowings.   
Many words are missing in \cite{D}  but
we have filled most of the gaps by finding the words on
Swadesh lists and on dictionaries freely available on the web.  
The file \cite{D} also contains information on ``cognates'' among
languages that we do not use in this work. 
Our selection of 50 languages is based on \cite{D} but 
we avoid to consider more than one version of the same language. 
For example, we do not consider both Irish A and
Irish B, but we choose only one of them and 
we do not include Brazilian but only Portuguese. 
Our choice among similar languages is based on keeping 
that language with fewer gaps.
Our database is available at \cite{footnote1}.

\subsection{Tree construction}
In this work the normalized Levenshtein is used to build up an
upper triangular $50 \times 50$  matrix with 1225 entries 
representing the pairwise distances corresponding to 50 languages. 
These genetic distances are translated into time distances between
language pairs and a new matrix of same size comes out. 
Then, the simple phylogenetic algorithm UPGMA \cite{UPGMA} 
for tree construction is used. 
The reason why we choose the unweighted pair-group method,
using arithmetic average (UPGMA), is that it is the 
most coherent with the hypothesis that the languages tree
is generated by  a coalescence process of
Kingman type \cite{K}. Let us describe briefly how this algorithm works. 
It first identifies the two languages with shortest time distance and then
it treats this pair as a new single object whose distance
from the other languages is the average of the distance of its two components.
Subsequently, among the new group of objects it identifies the pair with 
the shortest distance, and so on. 
At the end, one is left with only two objects (languages clusters)
which represents the two main branches at the root of the tree.
We remark that, as a consequence of the construction rule,
the distance between two branches is the average of the distances 
between all pairs of languages belonging to the two branches. 

\section{Conclusions}
We would like to compare now our results with those published in \cite{GA}.
The tree in Fig. \ref{fig1} is similar to the one in \cite{GA} 
but there are some important differences.
First of all, the first separation concerns Armenian, which forms a 
separate branch close to the root, while the other branch
contains all the remaining Indo-European languages.
Then,  the second one is that of Greek,
and only after there is a separation between the European branch and the
Indoiranian one.
This is the main difference with the tree in \cite{GA}, 
since therein the separation at the root
gives origin to two branches, one with Indoiranian languages 
plus Armenian and Greek, the other with European languages.
The position of Albanian is also different: in our case it is
linked to European languages while in \cite{GA} it goes with Indoiranian ones. 
Finally, the Romani language is correctly located together with 
Indian languages but it is not as close to Singhalese as reported in \cite{GA}.

In spite of this differences, our tree seems to confirm the 
same conclusions reported in \cite{GA} about the Anatolian origin of
the Indo-European languages, in fact, in our research, the first
separation concerns the languages geographically closer to Anatolia,
that is to say Armenian and Greek.

We want to stress that the method here used is very simple and does
not require any previous knowledge of languages origin. 
Also, it can be applied directly to all those language pairs for which
a translation of a small group of words exists. 
The results could be improved if more words are added to
the database and if translation and transliterations are
made more accurate.
Since our method is very easy to use, being the only difficulty the
procedure of collecting words, we plan to extend our study to other language families
and eventually test competing hypothesis concerning super families,
or test controversial classifications as for example the case of Japanese.

\section*{Acknowledgments}
We thank Julien Raboanary for many discussion and examples from Malagasy dialects
concerning the applicability of the Levenshtein distance to linguistics.
Critical  comments on many aspects of the paper by  M. Ausloos  
are also gratefully acknowledged.
The work by FP has been supported by European Commission Project
E2C2 FP6-2003-NEST-Path-012975  Extreme Events: Causes and Consequences.


\newpage

\newpage

\end{document}